\documentclass[aps,10pt,twocolumn,amsmath,amssymb,showpacs,floatfix,longbibliography]{revtex4-2}

\usepackage{graphicx}
\usepackage[utf8]{inputenc}

\usepackage[label font={large}]{subfig}
\usepackage[export]{adjustbox}
\usepackage{rotating}
\usepackage{booktabs, makecell, tabularx}
\usepackage{color}
\usepackage{dcolumn} 
\usepackage{bm}
\usepackage{mathtools}
\usepackage{latexsym}
\usepackage{float}
\usepackage{amsfonts}
\usepackage{amsmath}   
\usepackage{amssymb}  
\usepackage{babel}
\usepackage{natbib}
\usepackage[colorlinks]{hyperref}
\hypersetup{ 
	colorlinks=true, 
	linkcolor=blue, 
	filecolor=blue, 
	citecolor = blue,       
	urlcolor=blue, 
}

\DeclareUnicodeCharacter{2212}{-}

\begin{document}

\title{Screening of the Coulomb interaction in Carbon Nanotubes:\\ A First-Principles cRPA study}

\author{M. Rezayi}
\author{H. Hadipour}\email{hanifhadipour@gmail.com}

\affiliation{Department of Physics, University of Guilan, 41335-1914, Rasht, Iran \\}

\date{\today}

\begin{abstract}			
	We investigate the electronic screening of long-range Coulomb interactions in carbon nanotubes with different chiralities using first-principles calculations within the random-phase approximation. Depending on their wrapping vector, carbon nanotubes exhibit either metallic or semiconducting behavior, providing an ideal platform to explore how reduced dimensionality and electronic structure govern screening in one-dimensional systems. 
The strength of on-site Coulomb interactions in these compounds falls within the range of 3.5 to 5 eV, which is approximately 2$-$3 eV smaller than the corresponding values in nanoribbon compounds. This reduction subsequently affects the value of long-range interactions, consistent with experimental results regarding the smaller binding energy of excitons in nanotubes.  Despite their common carbon backbone, we find that the effective interaction landscape depends not only on metallicity but also sensitively on chirality and band topology. In particular, armchair and zigzag nanotubes with similar electronic character exhibit markedly different screening efficiencies. Our results establish a unified microscopic picture of electronic screening in carbon nanotubes and place them in direct context with previous first-principles studies of low-dimensional carbon nanostructures.   
\end{abstract}
		
\pacs{73.22.-f, 78.20.Bh, 71.15.-m, 71.10.Fd}

\maketitle

\section{Introduction} \label{sec:Introduction}

Carbon nanotubes (CNTs), cylindrical structures formed by rolling graphene sheets into seamless tubes, have long served as paradigmatic one-dimensional (1D) systems for exploring reduced-dimensional electronic phenomena \cite{RSC2009_CarbonNanotubeReview, Nano15151165_2025,lawal2025recent,Cao_2017}. Their electronic character depends sensitively on the chiral indices $(n,m)$, giving rise to either metallic or semiconducting behavior \cite{SWCNT_Electronic_Properties_PMC, Nano15151165_2025}. This tunability makes single-walled carbon nanotubes (SWCNTs) an ideal platform for investigating the interplay between dimensional confinement, electronic structure, and electron--electron interactions. In particular, the absence or presence of a band gap strongly influences transport, optical excitations, and collective response functions, as confirmed in recent experimental measurements showing that different chiralities yield markedly different carrier mobilities and on-state currents \cite{SWCNT_Chirality_NatComm_2023}.

In reduced-dimensional materials, Coulomb screening is intrinsically weakened compared to bulk solids, leading to pronounced many-body effects \cite{Deslippe2009_NanoLett}. In semiconducting CNTs, this is most prominently reflected in the formation of bound excitons that dominate their optical response \cite{Dukovic2005_NanoLett}. Resonant Raman and photoluminescence excitation experiments have measured exciton binding energies of $\sim\,0.49\pm0.05~\text{eV}$ and $\sim\,0.62\pm0.05~\text{eV}$ for specific semiconducting chiralities such as $(10,3)$ and $(7,5)$ SWCNTs, respectively \cite{Dukovic2005_NanoLett}, and complementary two-photon excitation spectroscopy across a range of semiconducting nanotubes reports binding energies in the range of $\sim0.3$--$0.4~\text{eV}$ \cite{Maultzsch2005_PRB}. More recent measurements on structure-sorted SWCNT assemblies yield exciton binding energies of approximately $0.26~\text{eV}$, demonstrating that moderate excitonic interactions \cite{Liu2025_ACSNano}. These binding energies originate from the reduced dielectric screening in quasi-one-dimensional systems, where electric field lines extend largely outside the nanotube, enhancing the effective electron--hole attraction \cite{Deslippe2009_NanoLett}.

In low-dimensional systems, reduced dimensionality strongly suppresses dielectric screening and leads to pronounced modifications of the effective Coulomb interaction, with important consequences for their electronic and correlated properties. First-principles studies have shown that in finite-size and quasi-one-dimensional systems, such as zero-dimensional clusters and atomically thin nanoribbons, the screened Coulomb interaction can deviate substantially from conventional bulk behavior and exhibit nonlocal and geometry-dependent screening \cite{Peters2,Hadipour2018,Montaghemi2020,Ramezani_2024}. In these systems, such unconventional screening is accompanied by significant many-body effects, including enhanced electron--hole interactions and large quasiparticle corrections, highlighting the central role of dimensionality in shaping correlation effects \cite{Amiri_2022,_a_o_lu_2017}. These findings motivate the study of how similar screening mechanisms appear in semiconducting CNTs, where quantum confinement and reduced dimensionality strongly affect Coulomb-driven phenomena.

Metallic CNTs, despite lacking an intrinsic band gap, exhibit nontrivial many-body response properties. Experimental magnetometry studies have revealed a pronounced anisotropy of the magnetic susceptibility in metallic SWCNTs, significantly exceeding that of their semiconducting counterparts \cite{Zaric2004}. This behavior has been attributed to strong orbital paramagnetism associated with gapless electronic states near the Fermi level \cite{Ajiki1993,Ando2004}. Furthermore, systematic measurements of intrinsic diamagnetic susceptibility demonstrate a clear dependence on tube diameter and electronic character \cite{Lu1995,Zaric2004}. These observations indicate that even in metallic CNTs, electron--electron interactions and screening effects play a decisive role in shaping their collective response.

Despite extensive experimental investigations of the optical and magnetic properties of CNTs, a systematic first-principles characterization of effective Coulomb interactions—particularly their spatial dependence and screening behavior across metallic and semiconducting nanotubes—remains incomplete. In particular, how reduced dimensionality and electronic character influence nonlocal and unconventional screening has not been explored in a unified framework. In this work, we employ \emph{ab initio} calculations within the random-phase approximation to evaluate static, spatially resolved effective Coulomb interactions in CNTs with different chiralities and electronic character. By placing our results in the context of experimentally observed excitonic phenomena, we provide a consistent microscopic picture of screening in one-dimensional carbon nanostructures.

\section{Computational Method}{\label{computational metod}}

\begin{figure}[!ht]
	\centering
	\includegraphics[scale=0.18]{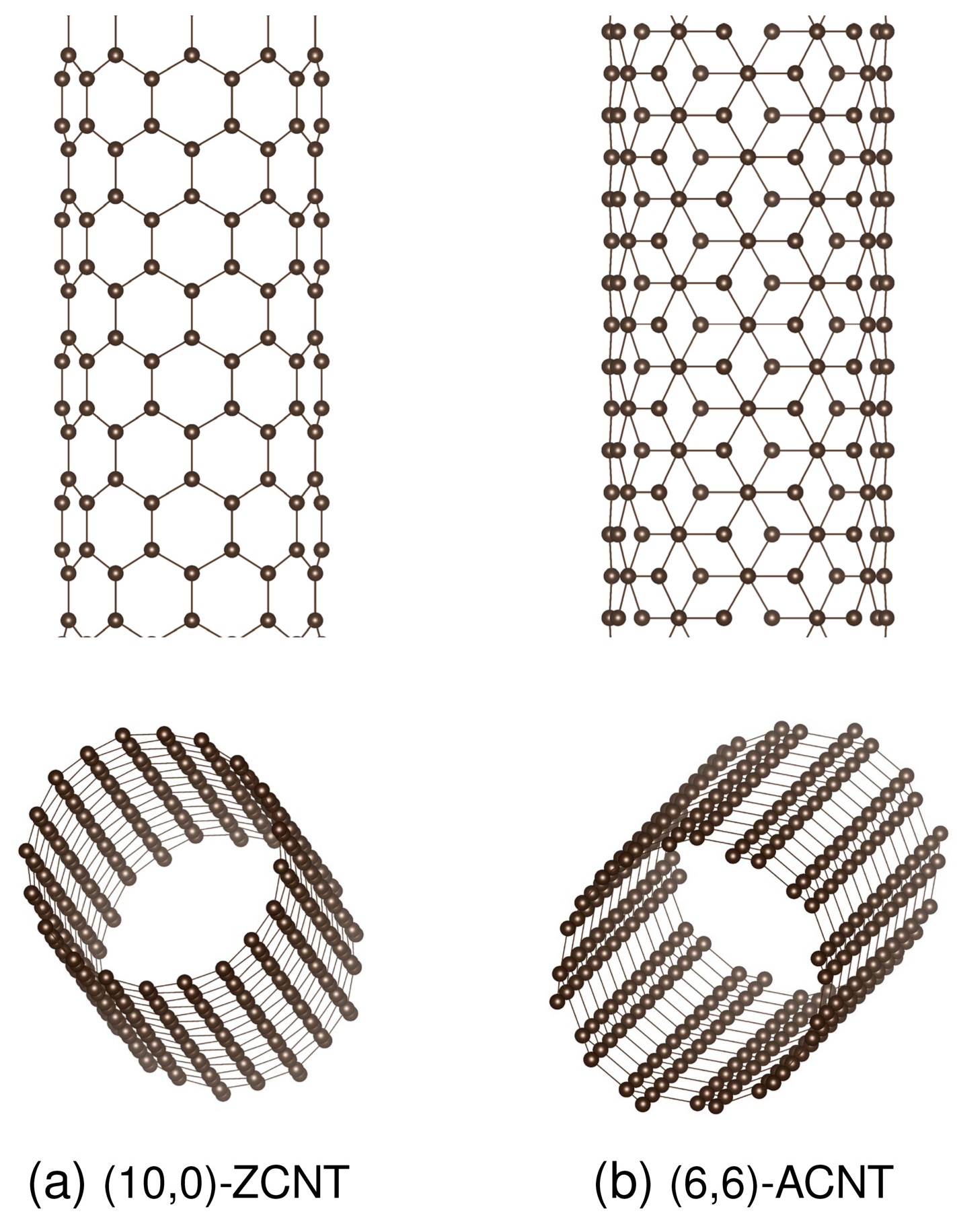}
	\caption{Carbon Nanotubes with two types of configuration: (a) Zigzag nanotube with $n$=10 [(10,0)-ZCNT], (b) Armchair nanotubes with $n$=6 [(6,6)-ACNT].}
	\label{fig:01}
\end{figure}

Generally, single-wall tubes can be characterized by two integers $(n,m)$ which specify the superlattice translation vector which wraps around the waist of the cylinder. Nanotubes of the type $(n,0)$ are called zigzag tubes, because they exhibit a zigzag pattern along the circumference. Such tubes display carbon-carbon bonds parallel to the nanotube axis. Also nanotubes of the type $(n,n)$ are called armchair tubes, because they exhibit an armchair pattern along the circumference. Such tubes display carbon-carbon bonds perpendicular to the nanotube axis. Fig.\ref{fig:01} shows some of these nanotubes.

Depending on the tube chirality, characterized by the indices $(n,m)$, two distinct electronic situations can occur. The first possible choice for $(n,m)$ nanotubes is given by the condition $ n-m=3l $. If two integers $(n,m)$ satisfied the condition $ n-m=3l $, the nanotubes are metallic, such as all armchair tubes and the $(n,0)$ zigzag tubes with n multiples of 3 are metallic nanotubes. The second possible choice for $(n,m)$ nanotubes is given by the condition $ n-m=3l\pm1 $. If two integers $(n,m)$ satisfied the condition $ n-m=3l\pm1 $, the nanotubes are semiconducting, such as the $(n,0)$ zigzag tubes with n exept the n that multiples of 3 are semiconducting nanotubes.

To quantify how reduced dimensionality and electronic character affect electron–electron interactions in these systems, we employ first-principles approaches to evaluate screened Coulomb interactions. This work investigates the partially and fully screened Coulomb interaction parameters computed using first-principles approaches: the constrained random-phase approximation (cRPA) and the conventional random-phase approximation (RPA), respectively~\cite{cRPA_1,cRPA_3,Vaugier,Karbalaee_Aghaee_2022,Yekta_2021,Hadipour_2019,Hadipour_2019x}.
The fully screened Coulomb interaction $W$ is related to the bare Coulomb
interaction $v$ as

\begin{equation}
W(\boldsymbol{r},\boldsymbol{r}',\omega)=\int d\boldsymbol{r}''  \epsilon^{-1}(\boldsymbol{r},\boldsymbol{r}'',\omega) v(\boldsymbol{r}'',\boldsymbol{r}'),
\label{eq1}
\end{equation}

where $\epsilon(\boldsymbol{r},\boldsymbol{r}'',\omega)$ is the dielectric function.

In the RPA of the dynamically screened Coulomb interaction, dielectric function  is related to
the electron polarizability $P$ by

\begin{equation}
\epsilon(\boldsymbol{r},\boldsymbol{r}',\omega)=\delta(\boldsymbol{r}-\boldsymbol{r}')-\int d\boldsymbol{r}'' v(\boldsymbol{r},\boldsymbol{r}'')P(\boldsymbol{r}'',\boldsymbol{r}',\omega),
\label{eq2}
\end{equation}
where the polarization function $P(\boldsymbol{r}'',\boldsymbol{r}',\omega)$ is given by
\begin{equation}
\begin{gathered}
P(\boldsymbol{r},\boldsymbol{r}',\omega)=\\
\sum_{\sigma} \sum_{\boldsymbol{k},m}^{occ} \sum_{\boldsymbol{k}',m'}^{unocc} \varphi_{\boldsymbol{k}m}^{\sigma}(\boldsymbol{r}) \varphi_{\boldsymbol{k}'m'}^{\sigma*}(\boldsymbol{r}) \varphi_{\boldsymbol{k}m}^{\sigma*}(\boldsymbol{r}') \varphi_{\boldsymbol{k}'m'}^{\sigma}(\boldsymbol{r}')  \\
\times\Bigg[ \frac{1}{\omega-\epsilon_{\boldsymbol{k}'m'}^{\sigma}-\epsilon_{\boldsymbol{k}m}^{\sigma}-i\eta} - \frac{1}{\omega+\epsilon_{\boldsymbol{k}'m'}^{\sigma}-\epsilon_{\boldsymbol{k}m}^{\sigma}-i\eta} \Bigg].
\end{gathered}
\label{eq3}
\end{equation}
\newline
Here $\epsilon_{\boldsymbol{k}m}^{\sigma}$ is single particle Kohn-Sham eigenvalues obtained from DFT and $\eta$ a positive infinitesimal.
Further, the $\varphi_{\boldsymbol{k}m}^{\sigma}(\boldsymbol{r})$ are the single particle Kohn-Sham eigenstates with
spin $\sigma$, wavenumber $\boldsymbol{k}$ and band index $m$. The tags $occ$ and $unocc$ above the summation symbol indicate that the
summation is respectively over occupied and unoccupied states only.

The cRPA method allows to calculate the effective Coulomb interaction (also called partially screened interaction) between
$p_z$ electrons in SWNTs. In this approach, in order to exclude the screening due to certain electrons we separates the full polarization
function in Eq.(\ref{eq3}) into two parts

\begin{equation}
P=P_{z}+P_{r},
\label{eq4}
\end{equation}
\newline
where $P_{z}$ includes only transitions between the $p_z$ states and $P_{r}$ is the remainder.
Then, the frequency dependent effective Coulomb interaction is given schematically by the matrix
equation
\begin{equation}
U(\omega) = [1-vP_{r}(\omega)]^{-1}v,
\label{eq5}
\end{equation}

\begin{figure}[t]
	\centering
	\includegraphics[scale=0.08]{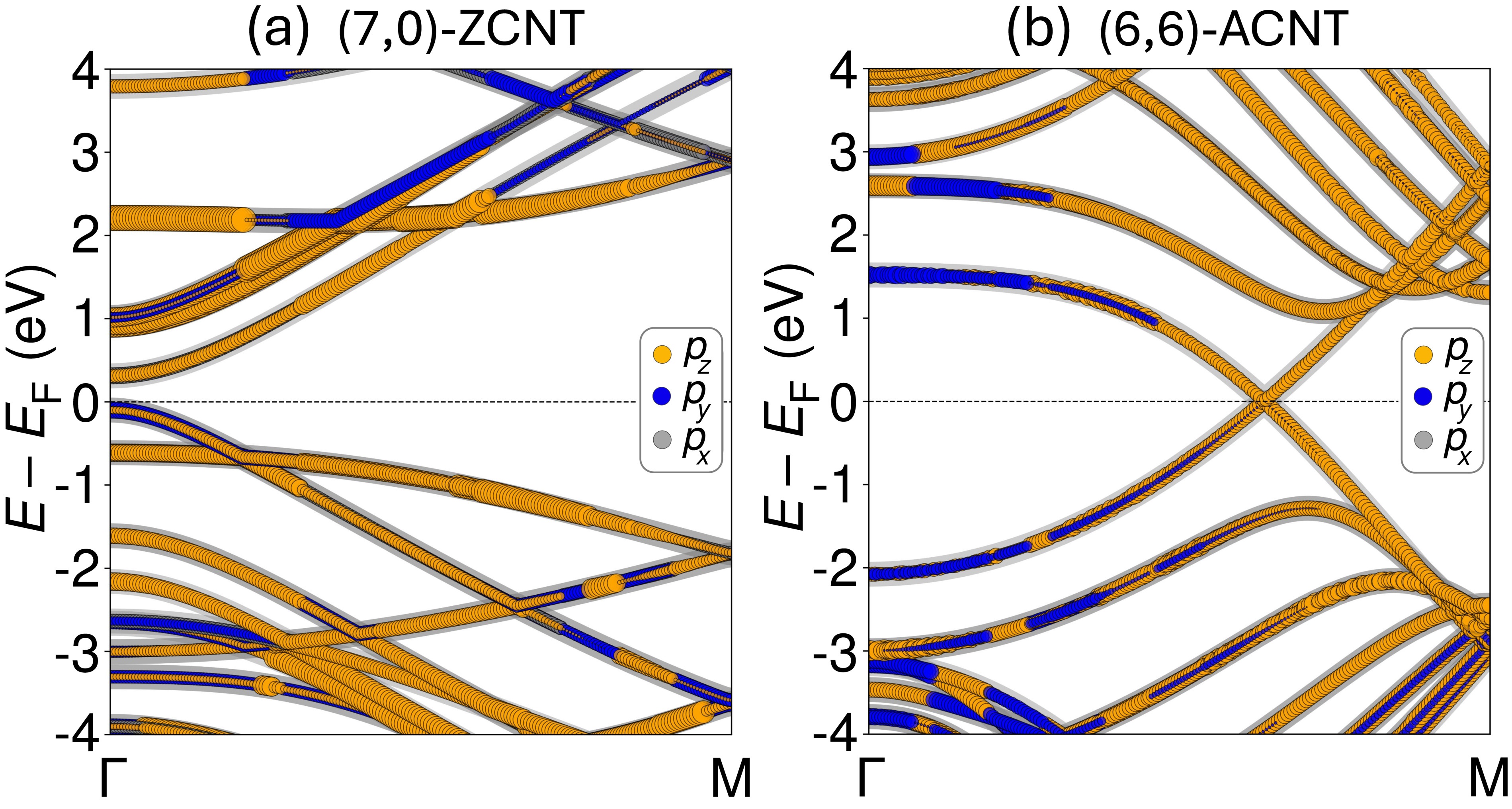}
	\caption{(a) Orbital-projected band structures for all C atoms of (a) (7,0)-ZCNT and (b) (6,6)-ACNT. The Fermi level is set to zero energy.}
	\label{fig:02}
\end{figure}

Application of  Eq.(\ref{eq4}) to materials with entangled bands is not straightforward.
Because $p_{z}$ states may be mixed with other extended $s$, $p_x$, and $p_y$ states at some parts
of the Brillouin zone, and thus  there is no unique identification of the $p_{z} \rightarrow p_{z}$
transitions for constructing $P_{z}$. To identify the correlated subspace and reveal the mixing of
$p_{z}$ states with other bands in the SWNTs, total and projected band structures for two systems
(7,0)-ZCNT and (6,6)-ACNT have been depicted in Fig.\ref{fig:02} (a) and (b) respectively.

For both systems the contribution of $p_{z}$ states around E$_{F}$ are significant as compared to $p_{x}$, $p_{y}$ , and $s$ orbitals.  The
$p_{z}$  bands are disentangled from the rest in large energy interval from -3 eV to 3 eV. For higher energies,
$p_{z}$ states are crossed by other bands and also mixed with them, which means they are entangled bands.
Several procedures have been proposed in the literature to calculate $P_{z}$ for entangled bands \cite{cRPA_1,cRPA_2}.
In this work we use the method described in Ref.\cite{cRPA_2}. Here we first define the probability to
find $p_{z}$ electrons of C atoms in eigenstate $\varphi_{\boldsymbol{k}m}^{\sigma}$ as,

\begin{equation}
c_{\boldsymbol{k}m}^{\sigma}=\sum_{i,n} \lvert T_{i,mn}^{\sigma \boldsymbol{k}} \rvert^{2},
\label{eq6}
\end{equation}
\newline
Here the unitary matrices $T_{i,mn}^{\sigma \boldsymbol{k}}$ are determined from the concept of maximally localized
Wannier functions,

\begin{equation}
w_{in}^{\sigma}(\boldsymbol{r})=\frac{1}{N} \sum_{\boldsymbol{k}}e^{-i\boldsymbol{k} \cdot \boldsymbol{R}_{i}} \sum_{m} T_{i,mn}^{\sigma \boldsymbol{k}} \varphi_{\boldsymbol{k}m}^{\sigma}(\boldsymbol{r}),
\label{eq7}
\end{equation}

where $w_{in}^{\sigma}(\boldsymbol{r})$ is a maximally localized Wannier function located at site $i$, $N$
is the number of discrete $k$ points in the full Brillouin zone and $\boldsymbol{R}_{i}$ the position vector
of atomic site $i$. The matrices $T_{i,mn}^{\sigma \boldsymbol{k}}$ are determined by minimizing the spread of
the Wannier functions,

\begin{equation}
\Omega=\sum_{i,n,\sigma}(\langle w_{in}^{\sigma} \lvert r^{2} \rvert w_{in}^{\sigma} \rangle - \langle w_{in}^{\sigma} \lvert \boldsymbol{r} \rvert w_{in}^{\sigma} \rangle^2).
\label{eq8}
\end{equation}

The summation in this context includes all Wannier functions. As evident from Eq. (\ref{eq7}), determining which bands to incorporate is crucial for constructing maximally localized Wannier states. The $p_{z}$ bands exhibit a broad energy range, overlapping with  $p_{x}$ and $p_{y}$ bands outside the -3 eV to 3 eV region. To ensure complete coverage of the correlated $p_{z}$ electron character, we have incorporated a large number of bands—specifically, 10 bands per carbon atom—in the Wannier orbital construction.

For entangled states the probability $c_{\boldsymbol{k}m}^{\sigma} < 1$ in Eq.(\ref{eq6}), while for
disentangled states $c_{\boldsymbol{k}m}^{\sigma}=1$. Then, the probability of an electron to be in the $p_z$
correlated subspace before and after a transition $\varphi_{\boldsymbol{k}m}^{\sigma} \rightarrow \varphi_{\boldsymbol{k}'m'}^{\sigma}$ is given by

\begin{figure}[b]
	\centering
	\includegraphics[scale=0.1]{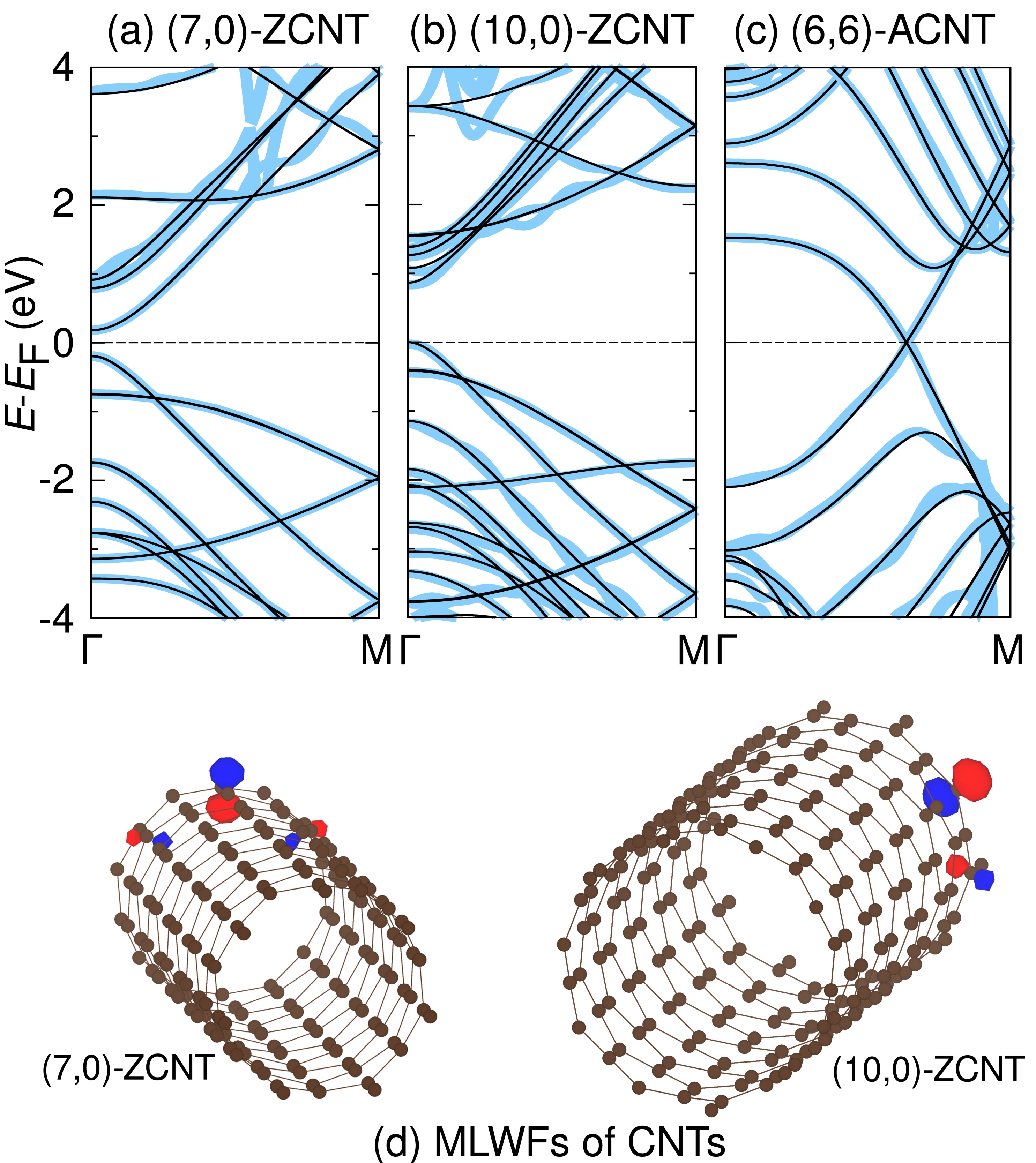}
	\caption{DFT-PBE and Wannier-interpolated band structure of  (a) (7,0)-ZCNT, (b) (10,0)-ZCNT and (c) (6,6)-ACNT. (d) Plot of $p_z$-like MLWFs for C atoms in (7,0)-ZCNT and (10,0)-ZCNT. Dashed lines denote the Fermi energy, which is set to zero.}
	\label{fig:03}
\end{figure}

\begin{equation}
p_{\boldsymbol{k}m\rightarrow\boldsymbol{k}'m'}^{\sigma}=c_{\boldsymbol{k}m}^{\sigma} c_{\boldsymbol{k}'m'}^{\sigma}.
\label{eq9}
\end{equation}
\newline
Thus for disentangled states $p_{\boldsymbol{k}m\rightarrow\boldsymbol{k}'m'}^{\sigma}=1$ and for entangled states $p_{\boldsymbol{k}m\rightarrow\boldsymbol{k}'m'}^{\sigma} < 1$. The polarization function $P_{z}$ now becomes

\begin{equation}
\begin{gathered}
P_{z}(\boldsymbol{r},\boldsymbol{r}',\omega)=\\
\sum_{\sigma} \sum_{\boldsymbol{k},m}^{occ} \sum_{\boldsymbol{k}',m'}^{unocc}(p_{\boldsymbol{k}m\rightarrow\boldsymbol{k}'m'}^{\sigma})^{2} \varphi_{\boldsymbol{k}m}^{\sigma}(\boldsymbol{r}) \varphi_{\boldsymbol{k}'m'}^{\sigma*}(\boldsymbol{r}) \varphi_{\boldsymbol{k}m}^{\sigma*}(\boldsymbol{r}') \varphi_{\boldsymbol{k}'m'}^{\sigma}(\boldsymbol{r}')  \\
\times\Bigg[ \frac{1}{\omega-\epsilon_{\boldsymbol{k}'m'}^{\sigma}-\epsilon_{\boldsymbol{k}m}^{\sigma}-i\eta} - \frac{1}{\omega+\epsilon_{\boldsymbol{k}'m'}^{\sigma}-\epsilon_{\boldsymbol{k}m}^{\sigma}-i\eta} \Bigg].
\end{gathered}
\label{eq10}
\end{equation}
\newline

The total polarization is calculated from Eq.(\ref{eq3}), and $P_{z}$ is obtained via Eq.(\ref{eq10}). $P_{r}$ can be
obtained from Eq.(\ref{eq4}).

For completeness, the effective Coulomb matrix within the selected subspace is computed by
\begin{equation}
\begin{gathered}
U_{in_{1},jn_{3},in_{2},jn_{4}}^{\sigma_{1},\sigma_{2}}(\omega)= \\
\int \int d\boldsymbol{r}d\boldsymbol{r}' w_{in_{1}}^{\sigma_{1}*}(\boldsymbol{r}) w_{jn_{3}}^{\sigma_{2}*}(\boldsymbol{r}') U(\boldsymbol{r},\boldsymbol{r}',\omega) w_{jn_{4}}^{\sigma_{2}}(\boldsymbol{r}') w_{in_{2}}^{\sigma_{1}}(\boldsymbol{r}).
\end{gathered}
\label{eq11}
\end{equation}
\newline

In the present case, we only consider a single Wannier orbital $w_R(r)$ of $p_z$ orbital character. The local
Hubbard $U$ parameter is then identical to the matrix element $U(\omega) =
\iint |w_{\mathbf{R}}(r)|^2 U(r,r';\omega) |w_{\mathbf{R}}(r')|^2 d^3r d^3r'$. Off-site elements
are defined as $U({\mathbf{R}}-{\mathbf{R'}},\omega)= \iint |w_{\mathbf{R}}(r)|^2 U(r,r';\omega) |w_{\mathbf{R'}}(r')|^2 d^3r
d^3r'$. In this study, we focus exclusively on the static limit ($\omega=0$).

Ground-state density functional theory (DFT) calculations are carried out using the FLAPW method
as implemented in the FLEUR code\cite{FLEUR}. We use the generalized gradient approximation
parameterized by Perdew \emph{et al.}\cite{Perdew} for the exchange correlation energy functional.
 $1\times1\times16$ \textbf{k}-point grids are used for unit cells of
SWNTs. A linear momentum cutoff of $G_{max}$ = 4.5 bohr$^{-1}$
is chosen for the plane waves. The DFT calculations are used as an input for the SPEX code~\cite{Schindlmayr}
to perform RPA and  constrained RPA calculations for the fully screened and partially screened (Hubbard $U$)
Coulomb interaction \cite{cRPA_1,cRPA_3,cRPA_4,Neroni_2019}.

To verify the validity of calculated Wannier functions, in Fig.\ref{fig:03}(a) to (c) we have presented
the DFT-PBE band structure and Wannier-interpolated bands obtained from the subspace selected by projecting onto
atomic $p_{z}$ orbitals on each C atom for (7,0)-ZCNT, (10,0)-ZCNT, and (6,6)-ACNT. As seen from the band structures, the overall agreement between original and Wannier-interpolated bands is quite good. Thus, in this context, the $p_{z}$ orbital
is the optimal correlated subspace for capturing the electronic
characteristics of these structures.

\begin{figure}[b]
	\centering
	\includegraphics[scale=0.09]{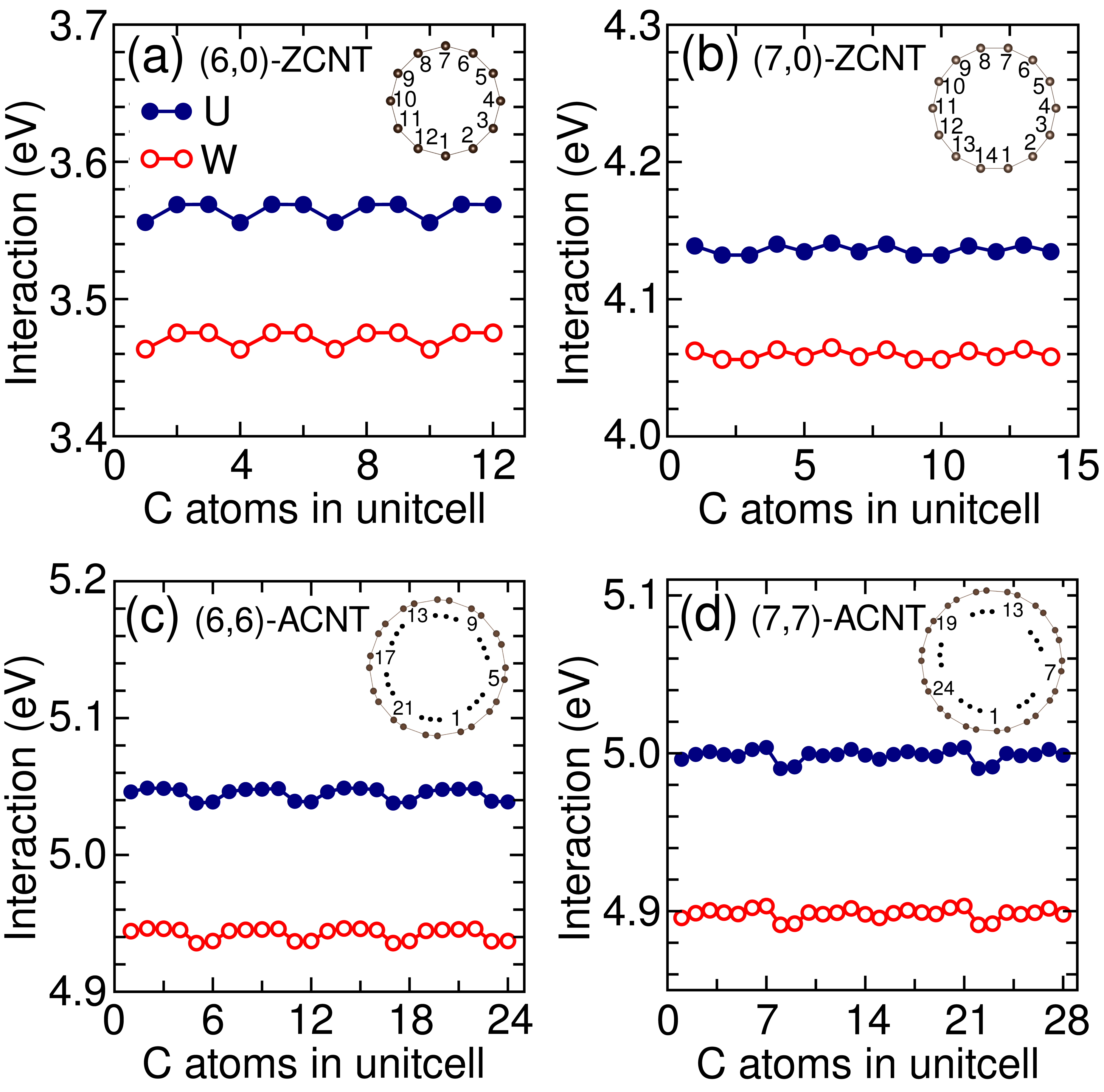}
	\caption{Calculated partially screened on-site interaction $U$ and
		fully screened Coulomb interaction $W$ parameters for carbon $p_z$
		electrons for four CNTs with armchair edges (a) (6,0)-ZCNT, (b) (7,0)-ZCNT, (c) (6,6)-ACNT, and (d) (7,7)-ACNT.}
	\label{fig:04}
\end{figure}

\section{Results and discussion}\label{sec3}

Our analysis begins by examining the on-site Coulomb interaction parameters in zigzag and armchair single-wall carbon nanotubes (ZCNTs and ACNTs) of four distinct diameters: $d_{a}$=4.7 $\AA$ for (6,0)-ZCNT, $d_{a}$=5.6 $\AA$ for (7,0)-ZCNT, and $d_{a}$=8.1 $\AA$ for (6,6)-ACNT, and $d_{a}$=9.4 $\AA$ for (7,7)-ACNT. Fig.\ref{fig:04} presents both the bare Hubbard $U$ and fully screened interaction $W$ parameters, revealing the role of $p_{z}$ electrons in the screening mechanism. The results are presented for two zigzag nanotubes, (6,0) and (7,0), and two armchair nanotubes, (6,6) and (7,7), covering both metallic and semiconducting electronic characters. For all considered systems, the values of $U$ and $W$ exhibit only weak variations among inequivalent carbon atoms within the unit cell, indicating a nearly homogeneous local screening environment and well-localized $p_z$-like Wannier functions.

Generally, the calculated Coulomb interaction in nanotube compounds is determined to be within the range of 3.5 to 5 eV. This value is notably smaller—by approximately 2-3 eV—than the corresponding interaction strength observed in nanoribbon structures \cite{Hadipour2018}. Consequently, the strength of long-range Coulomb interactions are also diminished. This overall enhancement in Coulomb screening is in excellent agreement with experimental observations pertaining to exciton binding energies in nanotubes, providing a consistent theoretical framework for understanding the optical properties of these materials.

A clear dependence of the on-site interaction strength on both chirality and electronic character emerges from Fig.~\ref{fig:04}. For the zigzag nanotubes, the semiconducting (7,0)-ZCNT exhibits a larger on-site interaction $U$ compared to the metallic (6,0)-ZCNT, reflecting the partial suppression of metallic screening channels upon opening a band gap. However, the increase of $U$ in the semiconducting zigzag nanotube remains moderate, and its absolute value is still smaller than that found in the armchair nanotubes. At the same time, the difference $U-W$ in (7,0)-ZCNT is relatively small, demonstrating that on-site screening is not fully quenched despite the semiconducting nature of this system.

This behavior highlights an important distinction between zigzag CNTs and quasi-one-dimensional nanoribbons studied previously. In contrast to semiconducting boron nitride  and graphene nanoribbons, where gap opening leads to a strong enhancement of the effective on-site interaction and a pronounced reduction of screening \cite{Hadipour2018,Montaghemi2020,Bagherpour2024}, the cylindrical geometry and electronic structure of zigzag CNTs allow for residual screening even in the presence of a band gap. The persistence of screening can be traced back to the substantial $p_z$ orbital weight distributed over the circumference of the nanotube, which facilitates nonlocal polarization processes.

For armchair nanotubes, both (6,6)- and (7,7)-ACNTs are metallic, yet their on-site interaction values are significantly larger than those of the zigzag systems. The calculated $U$ values are close to $5$~eV and remain nearly constant across different atomic sites. This demonstrates that metallicity alone does not uniquely determine the strength of the local Coulomb interaction. Instead, the topology of the electronic bands and the detailed orbital character near the Fermi energy play a decisive role. Similar conclusions were reached in earlier first-principles studies of graphene and phosphorene nanoribbons, where systems with comparable metallic character displayed markedly different effective interaction strengths depending on their edge geometry and orbital composition \cite{Hadipour2018,Bagherpour2024}.

Compared to pristine graphene, both $U$ and $W$ are reduced in all CNTs considered here, reflecting the modified screening environment induced by curvature and reduced dimensionality. This reduction is particularly pronounced in zigzag nanotubes, where typical on-site values of $U$ and $W$ are around $3.6$~eV and $3.4$~eV, respectively. Despite this overall reduction, a clear diameter dependence is observed, especially in armchair nanotubes, where both $U$ and $W$ exhibit noticeable oscillations as a function of tube diameter. These oscillations are more pronounced for the fully screened interaction $W$ and originate from quantum confinement effects that modulate the low-energy electronic structure and screening efficiency.

Taken together, these results show that opening a band gap in zigzag CNTs enhances the on-site Coulomb interaction but does not fully quench local screening, while armchair CNTs exhibit relatively large on-site interactions despite being metallic. The magnitude and trends of $U$ and $W$ are therefore governed by a subtle interplay of chirality, electronic structure, and orbital distribution, rather than by metallicity alone. In the following subsection, we connect these findings to the projected density of states to further elucidate the microscopic origin of the observed screening behavior.

\begin{figure}[!ht]
	\centering
	\includegraphics[scale=0.5]{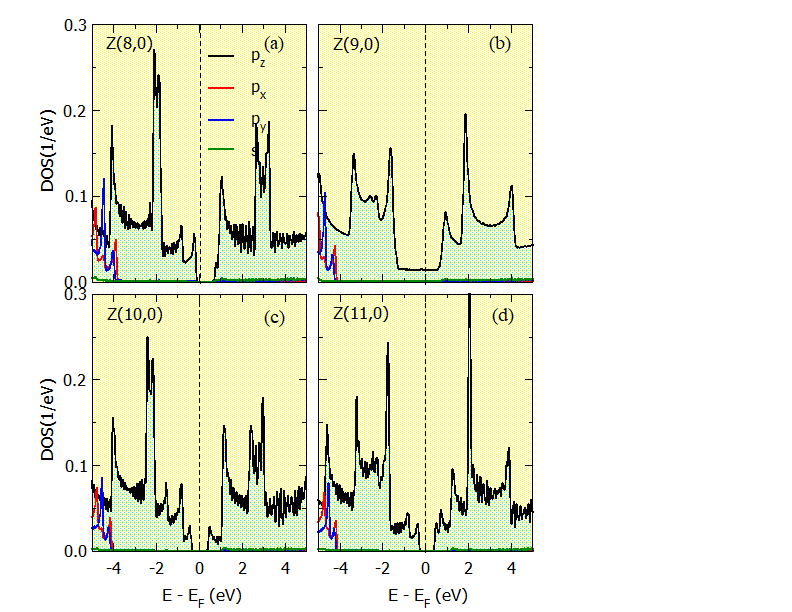}
	\caption{Projected density of state (PDOS) for some Zigzag Nanotubes such as (a) (8,0)-ZCNT, (b) (9,0)-ZCNT, (c) (10,0)-ZCNT, (d) (11,0)-ZCNT}
	\label{fig:05}
\end{figure}

\begin{figure}[!ht]
	\centering
	\includegraphics[scale=0.5]{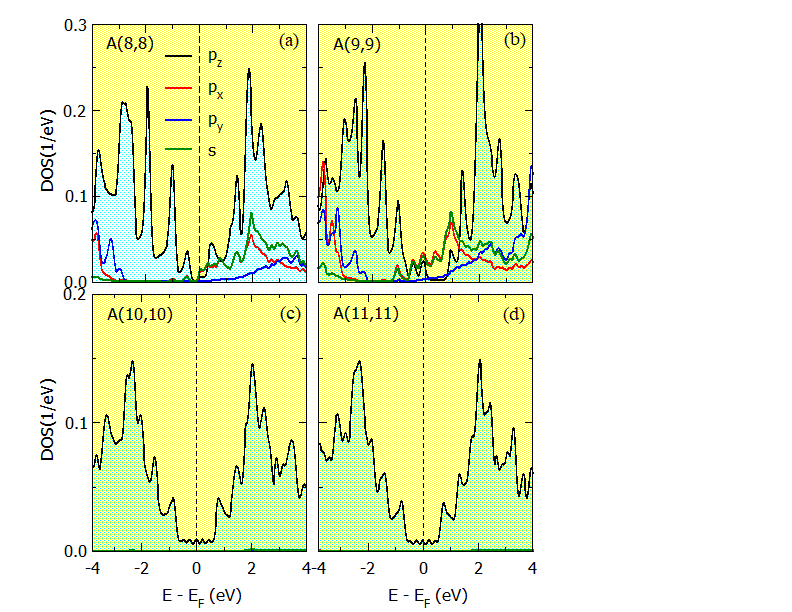}
	\caption{Projected density of state (PDOS) for some Armchair Nanotubes such as (a) (8,8)-ACNT, (b) (9,9)-ACNT, (c) (10,10)-ACNT, (d) (11,11)-ACNT}
	\label{fig:06}
\end{figure}

To elucidate the trends observed in the on-site Coulomb interaction parameters, we analyze the orbital-projected density of states (PDOS) for representative zigzag and armchair CNTs, as shown in Figs.~5 and 6. Although the PDOS data are presented for nanotubes with diameters slightly larger than those used in the Coulomb interaction calculations, their electronic structure is qualitatively representative and allows for a transparent interpretation of the screening behavior.

For zigzag nanotubes, Fig.~5 shows a clear distinction between metallic and semiconducting systems. In metallic ZCNTs, such as (9,0)-ZCNT, a substantial $p_z$-orbital weight is present at the Fermi level, providing low-energy particle--hole excitation channels that contribute efficiently to local screening. This explains the relatively smaller on-site interaction $U$ and the larger reduction from the bare interaction observed in metallic zigzag nanotubes. The delocalized nature of the $p_z$ states along the nanotube circumference further enhances the effectiveness of these screening processes.

In contrast, semiconducting zigzag nanotubes exhibit a depletion of the $p_z$ density of states at the Fermi level due to the opening of a band gap. Nevertheless, the PDOS reveals pronounced $p_z$ spectral weight in close proximity to the band edges. As a result, low-energy polarization channels are suppressed but not entirely eliminated. This partial reduction of screening leads to an increase of the on-site interaction $U$ compared to metallic zigzag nanotubes, while the difference $U-W$ remains relatively small. These findings indicate that the opening of a band gap in zigzag CNTs does not fully quench local screening, in agreement with the moderate enhancement of $U$ observed in Fig.~\ref{fig:04}.

A qualitatively different behavior is found for armchair nanotubes, as shown in Fig.~6. Although all considered ACNTs are metallic, their PDOS exhibits a broader and more structured $p_z$ contribution around the Fermi level, originating from multiple band crossings and the distinct topology of armchair bands. Despite the finite density of states at $E_F$, the screening of the on-site interaction is less effective than in metallic zigzag nanotubes. Consequently, the on-site Coulomb interaction $U$ in armchair CNTs is systematically larger. This demonstrates that metallicity alone does not uniquely determine the strength of local screening; instead, the detailed orbital character and band dispersion near the Fermi energy play a decisive role.

The observed PDOS trends provide a microscopic explanation for the behavior of the on-site Coulomb interaction parameters. In zigzag nanotubes, the transition from metallic to semiconducting character leads to a gradual reduction of screening rather than an abrupt suppression. In armchair nanotubes, the interplay between delocalized $p_z$ states and band topology results in comparatively weaker local screening despite the absence of a band gap. These features distinguish CNTs from quasi-one-dimensional nanoribbons, where edge-localized states and stronger quantum confinement lead to more pronounced screening suppression and larger effective interactions.

\begin{figure}[!ht]
	\centering
	\includegraphics[scale=0.15]{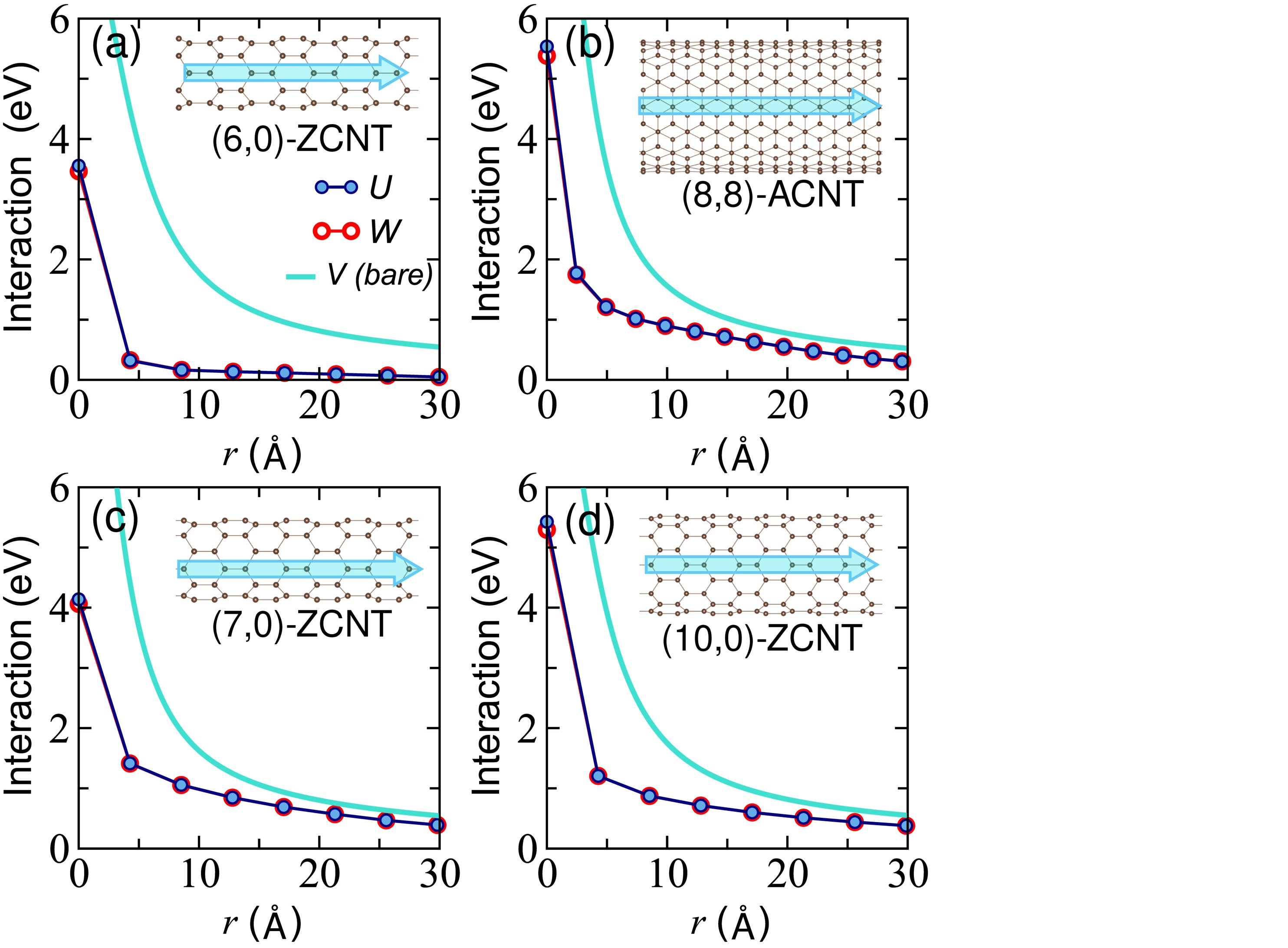}
	\caption{Distance dependence of the bare $V(r)$, the partially screened $U(r)$ obtained within cRPA, and the fully screened $W(r)$ calculated within RPA Coulomb interaction for representative zigzag and armchair CNTs, (a) (6,0)-ZCNT, (b) (8,8)-ACNT, (c) (7,0)-ZCNT, and (d) (10,0)-ZCNT. The interaction parameters are plotted for $p_z$ electrons as a function of the interatomic distance $r$ along the nanotube axis.}
	\label{fig:07}
\end{figure}

\begin{table*}[ht]
	\caption{On-site partially screened (Hubbard U) $U_{00}$, nearest-neighbor ($U_{01}$), next-nearest neighbor ($U_{02}$), third-nearest-neighbor ($U_{03}$), forth-nearest-neighbor ($U_{04}$), fifth-nearest-neighbor ($U_{05}$) and sixth-nearest-neighbor ($U_{06}$) Coulomb interaction parameters (in eV) for ZCNTs and ACNTs.}
	\begin{ruledtabular}
		\begin{tabular}{cclllllll}
			System & Ground State & \multicolumn{1}{c}{$U_{00}$} & \multicolumn{1}{c}{$U_{01}$} & \multicolumn{1}{c}{$U_{02}$} & \multicolumn{1}{c}{$U_{03}$} & \multicolumn{1}{c}{$U_{04}$} & \multicolumn{1}{c}{$U_{05}$} & \multicolumn{1}{c}{$U_{06}$} \\ \hline
			(6,0)-ZCNT & Metalic & 3.56 & 1.37 & 0.32 & 0.16 & 0.14 & 0.12 & 0.09 \\
			(7,0)-ZCNT & Semiconductor & 4.14 & 2.43 & 1.42 & 1.06 & 0.85 & 0.69 & 0.57 \\
			(8,0)-ZCNT & Semiconductor & 5.41 & 2.79 & 1.37 & 1.29 & 0.97 & 0.96 & 0.80 \\
			(9,0)-ZCNT & Metalic & 4.04 & 1.71 & 0.42 & 0.32 & 0.09 & 0.08 & 0.03 \\
			(10,0)-ZCNT & Semiconductor & 5.42 & 2.73 & 1.29 & 1.21 & 0.88 & 0.88 & 0.72 \\
			(6,6)-ACNT & Metalic & 5.05 & 2.06 & 1.46 & 1.25 & 1.08 & 0.92 & 0.76 \\
			(7,7)-ACNT & Metalic & 5.00 & 1.92 & 1.37 & 1.16 & 0.99 & 0.84 & 0.70 \\
			(8,8)-ACNT & Metalic & 5.54 & 1.77 & 1.21 & 1.02 & 0.90 & 0.81 & 0.72 \\
			(9,9)-ACNT & Metalic & 5.51 & 2.40 & 1.74 & 1.54 & 1.17 & 1.14 & 0.97
		\end{tabular}
		\label{table2x}
	\end{ruledtabular}
\end{table*}

While the on-site interaction already reveals important differences between chirality and electronic character, the reduced dimensionality of CNTs suggests that nonlocal Coulomb interactions may play an equally important role. We now turn to the spatial dependence of the effective Coulomb interaction and analyze the nonlocal matrix elements for both zigzag and armchair CNTs. Figure~7 shows the bare interaction $V(r)$, the partially screened interaction $U(r)$ obtained within cRPA, and the fully screened interaction $W(r)$ calculated within RPA as a function of the interatomic distance $r$. The corresponding numerical values for several neighbor shells are summarized in Table~I.

A clear qualitative distinction emerges between metallic and semiconducting zigzag nanotubes. In metallic ZCNTs, such as (6,0) and (9,0), the screened interaction decays rapidly with distance. Already beyond the nearest-neighbor separation, both $U(r)$ and $W(r)$ are strongly suppressed and approach negligible values at larger distances. This behavior reflects the presence of efficient low-energy screening channels associated with the finite density of states at the Fermi level, which effectively suppress long-range Coulomb interactions.

In contrast, semiconducting zigzag nanotubes, exemplified by (7,0)-ZCNT and (10,0)-ZCNT, exhibit a markedly different behavior. Although their on-site interaction is enhanced compared to metallic zigzag counterparts, the most striking feature is the persistence of sizable nonlocal interaction parameters over extended distances. As shown in Fig.~7(c) and (d), both $U(r)$ and $W(r)$ remain finite up to $r \sim 30$~\AA, indicating reduced screening efficiency in the absence of metallic polarization channels. This trend is quantitatively reflected in Table~I, where off-site interaction parameters up to the sixth-nearest neighbor remain substantial.

Despite this pronounced long-range character, we do not observe a regime in which the screened interaction exceeds the bare Coulomb interaction. Instead, the difference $V(r)-W(r)$ gradually decreases and approaches zero at large distances without changing sign. This behavior contrasts with previous first-principles studies on zero-dimensional clusters and quasi-one-dimensional nanoribbons, where nonconventional screening and antiscreening were reported \cite{Peters2,Hadipour2018,Montaghemi2020,Bagherpour2024}.

The absence of antiscreening in CNTs highlights an important geometrical effect. While reduced dimensionality generally weakens dielectric screening, the closed cylindrical geometry of nanotubes allows electric field lines to redistribute more symmetrically around the structure. This suppresses the dipolar imbalance responsible for antiscreening in open or finite geometries such as nanoribbons and molecular clusters. The persistence of weakened and spatially extended screening in semiconducting CNTs provides a natural microscopic context for the experimentally observed moderate excitonic effects in these systems, as reported in optical spectroscopy measurements \cite{Dukovic2005_NanoLett,Maultzsch2005_PRB,Liu2025_ACSNano}. Similar connections between reduced screening, nonlocal Coulomb interactions, and large exciton binding energies have been established previously for graphene, h-BN, and phosphorene nanoribbons, as well as for zero-dimensional clusters \cite{Peters2,Hadipour2018}.

Armchair nanotubes exhibit a distinct yet related behavior. Although all considered ACNTs are metallic, their nonlocal Coulomb interactions decay more slowly than in metallic zigzag nanotubes. As shown in Fig.~7(b) and Table~I, the effective interaction in armchair CNTs remains sizable over several neighbor shells, with appreciable values of $U_{02}$–$U_{06}$ even in the fully metallic regime. This demonstrates that metallicity alone is not sufficient to suppress nonlocal interactions. Although these systems are metals, the density of states near the Fermi level is very low, and they behave more like a semimetal. This can cause the Coulomb interaction to be large in armchair structures, similar to what happens in graphene.

Instead, the detailed band topology and the spatial distribution of low-energy $p_z$ states play a crucial role in determining screening efficiency. In metallic zigzag CNTs, the concentration of states near the Fermi level leads to strong local polarization channels and efficient short-range screening. In contrast, the more uniformly distributed electronic states (low DOS at $E_F$) in armchair CNTs give rise to less effective nonlocal screening, resulting in a relatively long-ranged effective interaction despite the absence of a band gap.

\section{Conclusions}
We have calculated the effective on-site and nonlocal Coulomb interaction parameters in metallic and semiconducting single-walled CNTs using a parameter-free first-principles approach based on the random-phase approximation and its constrained variant. The calculated interactions provide a microscopic description of screening in quasi-one-dimensional CNTs and are suitable for constructing realistic low-energy model Hamiltonians.
We find that the strength of Coulomb interaction  in nanotube compounds is in the range of 3.5$-$5 eV, which is roughly 2$-$3 eV smaller than that in nanoribbons. This reduction is not limited to short-range effects; it consistently extends to the long-range interaction terms as well. The resulting weaker overall Coulomb screening aligns perfectly with experimental measurements of exciton binding energies in nanotubes, validating our computational findings.

We find that opening a band gap in zigzag CNTs leads to an enhancement of the on-site Coulomb interaction; however, local screening is not fully suppressed and the increase of $U$ remains moderate compared to armchair nanotubes. In contrast, armchair nanotubes exhibit relatively large on-site interaction values despite being metallic, demonstrating that chirality and band topology play a decisive role in determining local Coulomb interactions beyond metallicity alone. The analysis of orbital-projected density of states shows that these trends are governed by the distribution and dispersion of low-energy $p_z$ states near the Fermi level.

Our investigation of nonlocal Coulomb interactions reveals a pronounced long-range character in semiconducting zigzag nanotubes, while metallic zigzag nanotubes display strongly short-ranged screening. Metallic armchair nanotubes show an intermediate behavior, with sizable off-site interaction parameters persisting over several neighbor shells. Despite the reduced screening and long-range nature of the effective interaction, we do not observe an antiscreening regime in CNTs: the difference between the bare and screened interactions approaches zero at large distances without changing sign.

The absence of antiscreening distinguishes CNTs from zero-dimensional clusters and quasi-one-dimensional nanoribbons and can be attributed to their relatively smaller non-local interactions and closed cylindrical geometry. Nevertheless, the weakened and spatially extended screening in semiconducting CNTs provides a consistent microscopic basis for the experimentally observed  excitonic effects in these systems. Our results establish nanotubes as a distinct class of low-dimensional materials in which nonlocal Coulomb interactions are moderate, qualitatively different from those in open or finite geometries.

\bibliography{my-bib}
\end{document}